\documentclass[twocolumn,secnumarabic,amssymb, nobibnotes, aps, prd]{revtex4}
\usepackage{fancybox}
\usepackage{color}
\usepackage{pstricks}

\usepackage{ifpdf}

\ifpdf
  \usepackage[pdftex]{graphicx}
\else
  \usepackage[dvips]{graphicx}
\fi

\begin{document}

\title{Hot Giant Loop Holography}%

\author{Gianluca Grignani$^1$, Joanna L.~Karczmarek$^2$ and Gordon
W.~Semenoff$^2$}
\affiliation{$^1$Dipartimento di Fisica, Universit\`a di Perugia,
INFN Sezione di Perugia, Via A. Pascoli, 06123 Perugia, Italia \\
$^2$Department of Physics and Astronomy, University of
British Columbia, Vancouver, British Columbia V6T 1Z1}

\begin{abstract}
We argue that there is a phase transition in the expectation value of
the Polyakov loop operator in the large N limit of the high
temperature deconfined phase of ${\cal N}=4$ Yang-Mills theory on a
spatial $S^3$.  It occurs for large completely symmetric
representation of the $SU(N)$ symmetry group. We speculate that this
transition is reflected in the D-branes which are the string theory duals
of giant loops.
\end{abstract}
\maketitle

The Hawking-Page phase transition \cite{Hawking:1982dh} is the
collapse of hot anti-de Sitter space to an anti-de
Sitter-Schwarzschild black hole.  A beautiful picture of the
holographic dual of this transition has emerged in the context of
AdS/CFT duality~\cite{Witten:1998zw}-\cite{Aharony:2003sx}.  On the
type IIB supergravity side is black hole formation with asymptotic
$AdS_5\times S^5$ geometry.  On the gauge theory side, we have the
deconfinement phase transition of large $N$ ${\cal N}=4$
supersymmetric Yang-Mills theory defined on the spatial three-sphere
$S^3$. The confined phase is dual to hot $AdS_5\times S^5$ whereas the
deconfined phase is dual to the black hole geometry.

Confining behavior of ${\cal N}=4$ Yang-Mills theory is governed by
the representation of its center symmetry and characterized by the
expectation value of the Polyakov loop, which is the trace of gauge
theory holonomy in periodic Euclidean
time~\cite{Polyakov:1978vu}\cite{Susskind:1979up},
\begin{equation}\label{holonomy}
{\rm Tr}~ U(\vec x) ={\rm Tr}~{\cal P}e^{i\int_0^\beta d\tau
A_0(\tau,\vec x)}~.\end{equation} The loop transforms as ${\rm
Tr}~U(\vec x)\to c {\rm Tr}~U(\vec x)$, where $c=e^{2\pi i/N}$ is the
generator of the $Z_N$ center of the $SU(N)$ gauge group. Its
expectation value vanishes in the low temperature confining phase
where center symmetry is good, and it can be non-zero in the high
temperature deconfined phase where the center symmetry is
broken.

The study of the Polyakov loop using effective field theory has a long
history \cite{McLerran:1980pk}; however, it was applied to ${\cal
N}=4$ Yang-Mills theory on a spatial $S^3$ only relatively
recently~\cite{Sundborg:1999ue}\cite{Aharony:2003sx}.  In the weak
coupling limit, all of the fields in ${\cal N}=4$ Yang-Mills theory
have a mass gap arising from their conformal coupling to the curvature
of the $S^3$.  Integrating them out to find an effective field theory
in which one can compute the expectation value of the Polyakov loop is
a well-defined procedure.  The effective theory is a unitary
one-matrix model with effective action $S_{\rm eff}[U]$. Gauge
symmetry implies $S_{\rm eff}[U]=S_{\rm eff}[WUW^\dagger]$, where $W$
is a unitary matrix, while center symmetry implies $S_{\rm
eff}[U]=S_{\rm eff}[c U]$.  Further, the action is of order $N^2$,
$S_{\rm eff}[U=1]\sim N^2$.  Since ${\cal N}=4$ Yang-Mills theory has
conformal symmetry, the effective action depends on the temperature
$T$ and the $S^3$ radius $R$ through the product $TR$. It also depends
on the 't Hooft coupling $\lambda=g_{YM}^2N$ and on $N$ which we
assume is taken to infinity holding $\lambda$ fixed.  At weak
coupling, where $\lambda\to0$, the phase transition is found by tuning
$TR$ to a critical value.  When the coupling $\lambda$ is turned on,
this phase transition is thought to persist and at large $\lambda$ to 
coincide with the Hawking-Page transition of the gravity dual.  The
effective field theory description should be reliable when $TR<1$;
however, it is thought to have a broader applicability.  We will
assume that it can be used to discuss the deconfined phase, at least
in the vicinity of the phase transition that occurs when $TR\sim 1$.

The unitary matrix model can be used to calculate the expectation
value of the Polyakov loop operator in any irreducible representation
$R$ of the $SU(N)$ gauge group,
\begin{equation}
\left< {\rm Tr}_{R}U(x) \right>=
\frac{\int [dU] e^{-S_{\rm eff}[U]} {\rm Tr}_R~U
} {\int [dU] e^{-S_{\rm eff}[U]} }~.
\label{ploop}
\end{equation}
We shall show in the following that (\ref{ploop}) can have interesting
behavior which depends on the size and nature of the representation.
We will consider completely symmetric representations ${\cal S}_k$
whose Young tableau is a single row with $k$ boxes and completely
antisymmetric representations ${\cal A}_k$ whose Young tableau is a
single column with $k$ boxes. We shall
consider large values of $k$ so that $\frac{k}{N}$ remains finite as
$N\to\infty$.
Both representations ${\cal S}_k$ and ${\cal A}_k$ have center charge
$k$ mod $N$ so that (\ref{ploop}) vanishes in the confined phase when this
charge is non-zero.  The characters can be non-zero in the deconfined phase.

In the duality between gauge fields and strings, the expectation value
of the Wilson loop is dual to an open fundamental string amplitude.
This has been made precise for the Maldacena-Wilson
loop~\cite{Maldacena:1998im} which differs from the Polyakov loop
(\ref{ploop}) in that it contains the scalar fields of the ${\cal
N}=4$ theory as well as the gauge field. In that case, the boundary of
the fundamental string worldsheet is located on the loop contour
placed at the asymptotic boundary of $AdS_5$.  In the zero temperature
Yang-Mills theory defined on a spatial $R^3$, an interesting
phenomenon occurs for loops in representations where the number of
boxes $k$ in the Young tableau is large so that $\frac{k}{N}$ is finite
in the large $N$ limit. The dual fundamental string worldsheet is
replaced by a D-brane with world-volume electric
flux~\cite{Drukker:2005kx}-\cite{RodriguezGomez:2006zz}.  This was
found by studying highly supersymmetric $\frac{1}{2}$-BPS loops, where
some results are known for all values of the coupling constant
\cite{Erickson:2000af}. For the anti-symmetric representation, the
dual is a D5-brane whose world volume is a direct product of
$AdS_2\subset AdS_5$ and $S^3\subset S^5$. For a symmetric
representation, it is a D3-brane with world volume $AdS_2\times
S^2\subset AdS_5$.  It is interesting to ask whether these D-branes
exist in the finite temperature geometry where they would be dual to a
gauge theory loop linking periodic Euclidean time.  This question has
already been studied by Hartnoll and Kumar \cite{Hartnoll:2006hr} who
looked for solutions of the appropriate Born-Infeld actions on the
black hole background.  For the D5-brane wrapped on $S^4\subset S^5$
which corresponds to a totally antisymmetric representation on the
gauge theory side, there seem to be solutions for any $\frac{k}{N}$
with the usual cutoff at $k=N$ dictated by the maximum size of an
antisymmetric representation on the gauge theory side and a maximum
radius for embedding $S^4$ in $S^5$ on the supergravity side.
However, in the case of the D3-brane which should correspond to a
totally symmetric representation, Hartnoll and Kumar could not find
any solutions at all. This fundamental difference between the two
cases is what motivated our work on the gauge theory which we shall
now summarize.  Afterward, we will revisit the question of the
supergravity side. We emphasize that in supergravity we are studying
the dual of the Maldacena-Wilson loop whereas in gauge theory our
analysis is limited to the Polyakov loop. Both are governed by the
center symmetry and both can become non-zero at the
decnfinement transition.  At high temperature, due to decoupling of
the scalar fields, they should become similar.

To study large totally symmetric or totally antisymmetric
representations, it is convenient to obtain the characters in
(\ref{ploop}) from generating functions,
\begin{equation}\label{sym}
e^{-N\Gamma_{{\cal S}_k/{\cal A}_k}}\equiv
\langle {\rm Tr}_{{\cal S}_k/{\cal A}_k}~U \rangle
= \oint dt
\frac{\langle e^{\mp{\rm Tr}\ln(1 \mp tU)}\rangle}{2\pi i t^{k+1}}~,
\end{equation}
where the upper/lower sign is for symmetric/antisymmetric
representations (${\cal S}_k/{\cal A}_k$) respectively and the contour
in the integral over $t$ encircles the origin.  In the large $N$
limit, these integrals can be computed using two saddle point
approximations.  The first occurs while integrating over unitary
matrices in (\ref{ploop}).  Because of the gauge symmetry, this is an
eigenvalue model -- the gauge symmetry can be used to diagonalize
$U={\rm diag}[e^{i\phi_1},...,e^{i\phi_N}]$. At large $N$, the
eigenvalues become classical variables and their distribution is found
by minimizing $S_{\rm eff}$ plus a Jacobian from the unitary integral
measure.  As long as $k<<N^2$, the loop operators in (\ref{sym}) do
not modify the eigenvalue distribution which is given by a density
$\rho(\phi)$. $\rho(\phi)d\phi$ is $\frac{1}{N}$ times the number of
eigenvalues between $\phi$ and $\phi+d\phi$ and is normalized,
$\int_{-\pi}^\pi d\phi\rho(\phi)=1$.  Center symmetry is now an
invariance under a simultaneous translation of all eigenvalues,
$\phi_a\to \phi_a+$constant. In the center-symmetric confined phase,
the distribution is translation invariant, eigenvalues are uniformly
distributed on the unit circle and $\rho_{\rm conf}=\frac{1}{2\pi}$. In
the de-confined phase, the eigenvalues are clumped. We will assume
their distribution is symmetric about zero so that
$\rho(\phi)=\rho(-\phi)$. In the large $N$ limit the expectation
values in Eq.~(\ref{sym}) are computed using the eigenvalue density,
\begin{equation}\label{sym2}
e^{-N\Gamma_{{\cal S}_k/{\cal A}_k}}
=
\oint dt
\frac{e^{\mp N\int_{-\pi}^\pi d\phi\rho(\phi)\ln(1\mp te^{i\phi})} }{2\pi i t^{k+1}}~.
\end{equation}
The second use of a saddle-point approximation is to evaluate the
integral over $t$ in (\ref{sym2}).  Let
$\hat t$ satisfy the saddle-point equation
\begin{equation}\label{sym4}
R_{{\cal S}_k/{\cal A}_k}(\hat t) \equiv\int_{-\pi}^\pi d\phi\rho(\phi) \frac{\hat
te^{i\phi}}{1\mp\hat te^{i\phi}}=\frac{k}{N}~.
\end{equation}
Then, the free energy is given by
\begin{equation}\label{sym3}
\Gamma_{{\cal S}_k/{\cal A}_k}=\pm\int^{\pi}_{-\pi} d\phi \rho(\phi)
\ln(1\mp\hat te^{i\phi})+\frac{k}{N}\ln \hat t~.
\end{equation}
The functions $R_{{\cal S}_k/{\cal A}_k}(t)$ in (\ref{sym4}) are
related to the resolvent of the matrix model and are holomorphic
functions of $t$ with cut singularities on the unit circle determined
by the support of $\rho(\phi)$.

Let us begin with the symmetric representation ${\cal S}_k$.  We
shall consider three examples of eigenvalue distributions. First, the
confining phase has $\rho_{\rm conf}= \frac{1}{2\pi}$. $R_{{\cal
S}_k}(t)$ vanishes if $|t|<1$ and is $-1$ if $|t|>1$.  This is the
expected discontinuity at the unit circle.  Eq.~(\ref{sym4})
has solutions only when $\frac{k}{N}=0$, consistent with confinement.

As a second example consider $ \rho(\phi) =\frac{1}{2\pi}\left(
1+2p\cos\phi\right) $.  $p=\frac{1}{N}\langle{\rm Tr}~U\rangle=\int
d\phi\rho(\phi)e^{i\phi}$ is the fundamental representation
loop. Positivity of the density requires $0\leq p\leq \frac{1}{2}$.
This distribution depends on $\phi$ and therefore is deconfined. While
it is not realistic for ${\cal N}=4$ Yang-Mills theory, it does occur
in the strong-coupling phase of large $N$ 2-dimensional lattice
Yang-Mills theory~\cite{Gross:1980he}.

There is one solution of $R_{{\cal S}_k}(\hat t)=\frac{k}{N}$ in the
region $|\hat t|<1$ at $\hat t=\frac{k}{N}/p$. (If $frac{k}{N}$ and
$p$ are such that $|\hat t|>1$, both $R_{{\cal S}_k}$ and
$\Gamma_{{\cal S}_k}$ should be extended there by analytic
continuation.)  The free energy is
\begin{equation} \Gamma_{{\cal
S}_k}=\frac{k}{N}\ln \left[\frac{k/N}{ep}\right]~,
\label{symfe}\end{equation}
where $e=2.718\ldots$.
$\Gamma_{{\cal S}_k}$ has the interesting feature that, as
$\frac{k}{N}$ is increased, it changes sign from negative to positive.
This results in a {\bf phase transition}, which occurs when
$\frac{k}{N}=\left(\frac{k}{N}\right)_{\rm crit}=ep$.  When
$\frac{k}{N}<\left(\frac{k}{N}\right)_{\rm crit}$, $\Gamma_{{\cal S}_k}$ is
negative and the loop expectation value, $e^{-N\Gamma}$, is
exponentially large.  When $\frac{k}{N}>\left(\frac{k}{N}\right)_{\rm crit}$,
$\Gamma_{{\cal S}_k}$ is positive and the loop vanishes for
$N\to\infty$. This phase transition implies that, even in the
deconfined phase, sufficiently large symmetric representations are
still confined.

At this point, the reader might wonder how the expectation value of a
unitary matrix can grow exponentially.  The exponential comes from the
traces needed to get the large representation and which give $k$ and
$N$ dependent factors.  Note that we did not normalize the loop (which
would divide by a $k$ and $N$-dependent factor).  In
Ref.-\cite{Gomis:2006sb}, it was shown that it is the un-normalized
loop that should be compared with supergravity, which is our eventual
aim.

As a check of the saddle-point approximation to the $t$-integral in
this simple example, observe that, if for the moment we assume that
$k$ and $N$ are finite, we can integrate (\ref{sym2}) explicitly to
get $e^{-N\Gamma_{{\cal S}_k}}= \frac{N^k}{k!}p^k$. Using the Stirling
formula and taking $k\sim N\to\infty$ reproduces (\ref{symfe}).

We note that the presence of the phase transition is a universal property
of the confining phase.
From (\ref{sym3}) and (\ref{sym4}), we see that
$\frac{d\Gamma_{{\cal S}_k}}{d(k/N)}=\ln\hat t$ where $\hat t$ solves
(\ref{sym4}).
Further, by inspecting (\ref{sym3}) we see that $\Gamma_{{\cal S}_k}$
is real and negative when $\hat t$ is real and $\hat t<1$. As $\hat t$
increases, $\Gamma_{{\cal S}_k}$ decreases to a minimum at
$\hat t=1$, then begins increasing in the region $\hat t>1$ and
eventually becomes positive.  $\hat t$ increases with $\frac{k}{N}$
throughout this region.

To see this behavior in another example, consider the semi-circle
distribution which, for $|\phi|<2\arcsin\sqrt{2-2p}$, is
\begin{equation}\label{semicircle}
\rho(\phi)=\frac{\cos\frac{\phi}{2}}{\pi(2-2p)}\sqrt{2-2p-
\sin^2\frac{\phi}{2}}
\end{equation}
and which vanishes in the gap $2\arcsin\sqrt{2-2p}\leq |\phi|\leq\pi$.  We
still use the fundamental loop, $p$, as a parameter and now
$\frac{1}{2}\leq p\leq 1$.  This is the distribution in the weak
coupling phase of 2-dimensional lattice Yang-Mills theory
\cite{Gross:1980he}.  It is also an approximation to the deconfined
distribution for weakly coupled ${\cal N}=4$ Yang-Mills theory
\cite{Aharony:2003sx}\cite{Jurkiewicz:1982iz}. For sufficiently
weak coupling, it could be accurate near the phase transition where
$p=\frac{1}{2}$.  The saddle point computation can be done explicitly
near t=0 and analytically continued.
The free energy is
\begin{eqnarray}\label{semisymmfreeenergy1}
\Gamma_{{\cal S}_k}&=&
\left(2\theta\cosh\theta-\sinh\theta\right)
\frac{ \sinh\theta+\sqrt{\sinh^2\theta+2-2p}}{2-2p}
\nonumber \\ \label{semisymmfreeenergy2}
&-&\frac{1}{2}
-\ln\left[\frac{ \sinh\theta+\sqrt{\sinh^2\theta+2-2p}}
{2-2p}\right]~,
\end{eqnarray}
where $\theta$ is defined by $\hat t=e^{2\theta}$ and is
determined by the saddle-point equation
\begin{equation}\label{saddlepointequation}
\frac{k}{N}+\frac{1}{2}=\cosh\theta\left[\frac{\sinh\theta+
\sqrt{\sinh^2\theta+2-2p}}{2-2p}\right]~,
\end{equation}
which can be solved for $\sinh(\theta)$.  The free energy is zero when
$k=0$, negative for small $k$, goes to zero at a critical
$\frac{k}{N}$ and is positive thereafter. This is so for any value of
$p$ in the allowed range.  A graph of $\Gamma_{{\cal S}_k}$
versus $\frac{k}{N}$ for $p=0.51$ is plotted in Fig.~1. With this
value of $p$, the free energy becomes positive at $\theta\simeq 0.50$
which corresponds to $\frac{k}{N}_{\rm crit.}\simeq 1.3$.
\begin{figure}
\includegraphics[scale=0.38]{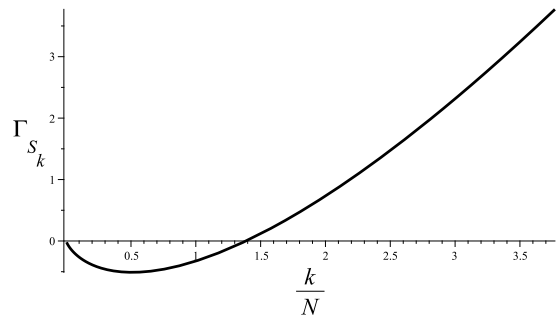}
\caption  {The free energy $\Gamma_{{\cal S}_k}(\theta)$ as a function of $\frac{k}{N}$
in the semi-circle distribution with $p=0.51$.}\end{figure}

Now, we consider the antisymmetric representation. For a large class
of distributions gapped around $\phi=\pi$ and with $\frac{d R_{{\cal
A}_k}(\hat t)} {d \hat t} > 0$, which includes the semi-circle
distribution (\ref{semicircle}), we can argue that $\Gamma_{{\cal
A}_k}$ is always negative and the phase transition that we are
discussing does not occur. To begin, by changing variables in
(\ref{sym3}), we observe that $\Gamma_{{\cal A}_k} =\Gamma_{{\cal
A}_{N-k}}$. This symmetry is reflected in the saddle-point equation
(\ref{sym4}) which, using our assumption that
$\rho(\phi)=\rho(-\phi)$, can be re-written as
\begin{eqnarray}\label{cut}
\frac{1}{2}\int_{-\pi}^\pi d\phi\rho(\phi)
\frac{\hat t^{\frac{1}{2}}e^{i\frac{\phi}{2}}-\hat t^{-\frac{1}{2}}e^{-i\frac{\phi}{2}}}
{\hat t^{\frac{1}{2}}e^{i\frac{\phi}{2}}
+\hat t^{-\frac{1}{2}}e^{-i\frac{\phi}{2}}}
=\frac{k}{N}-\frac{1}{2}
\end{eqnarray}
and implies $\hat t(k/N)=1/\hat t(1-k/N)$.
The free energy,
{\small\begin{equation}\label{sym6}
\Gamma_{{\cal A}_k}
=-\int_{-\pi}^\pi d\phi\rho(\phi)
\ln\left(\hat t^{\frac{1}{2}}e^{i\frac{\phi}{2}}+\hat
t^{-\frac{1}{2}}e^{-i\frac{\phi}{2}}\right)
+\left(\frac{k}{N}-\frac{1}{2}\right)\ln \hat t
\end{equation} }
\hspace{ -.08 cm}is symmetric under $\frac{k}{N}\to
1-\frac{k}{N}$. Moreover, with a gapped distribution, $\rho(\pi)=0$,
and the integral in (\ref{cut}) is continuous at $\hat t=1$. From
{\small $\frac{d R_{{\cal A}_k}(\hat t)} {d \hat t} > 0$}, $\hat t(k)$
is monotone, and one can see in (\ref{cut}) that $\hat t=0$
corresponds to $\frac{k}{N}=0$, $\hat t=\infty$ to $\frac{k}{N}=1$ and
$\hat t=1$ to $\frac{k}{N}=\frac{1}{2}$.  Furthermore, since
$\frac{d\Gamma_{{\cal A}_k}}{d(k/N)} =\ln \hat t(k)$,
$\frac{d^2\Gamma_{{\cal A}_k}}{d(k/N)^2}>0 $, thus $\Gamma_{{\cal
A}_k}$ is a convex function which decreases from $0$ to a negative
minimum as $\frac{k}{N}$ goes from $0$ to $\frac{1}{2}$ and then
increases back to zero when $\frac{k}{N}$ goes from $\frac{1}{2}$ to
$1$.  $\Gamma_{{\cal A}_k}$ does not become positive and there is no
phase transition of the kind that we found for symmetric
representations. When the distribution is ungapped, or when $\frac{d
R_{{\cal A}_k}(\hat t)} {d \hat t}$ becomes negative (for example when
$p
< 0$), interesting behavior can occur.  We put off a discussion of it
to a future investigation.

We have found a difference between the symmetric and antisymmetric
representation Polyakov loops which is qualitatively similar to the
one found by Hartnoll and Kumar~\cite{Hartnoll:2006hr} for the dual
objects in supergravity: the antisymmetric loop is non-zero in the
deconfined phase for all allowed $\frac{k}{N}$ and the dual D5-brane
exists whereas the gauge theory symmetric representation loop has a
phase transition.  The numerical search for the dual D3-brane in
\cite{Hartnoll:2006hr} combined with analytic arguments at large
$\kappa={\sqrt \lambda\over 4}{k\over N}$ found no solution.  
If we take this to mean that the expectation value of
the gauge theory quantity vanishes, it suggests that, at strong
coupling, the critical value $\frac{k}{N}$ goes to zero faster than
$\frac{4}{\sqrt{\lambda}}$.

We can also examine the alternative that the phase transion occurs for
a value of $\kappa$ so small that solutions are missed in the
numerical analysis.  The gravity background is the asymptotically
$AdS_5\times S^5$ black hole metric 
{\small\begin{equation}
dS^2=R^2\left[f(r)dt^2+\frac{dr^2}{f(r)}+r^2\left(d\chi^2+ \sin^2\chi
d\Omega_2^2\right)+d\Omega_5^2\right]~, \label{metric}
\end{equation}}
where $ f(r)= (r^2-r_+^2)(r^2+1+r_+^2)/r^2 $.  The horizon is located
at $r=r_+$ and time is identified periodically with period $\frac{2\pi
r_+}{2r_+^2+1}$, the inverse Hawking temperature of the black hole.
The D3-brane world volume embedded in this geometry is {\small
\begin{equation}
\label{geometry} ds^2=R^2\left[f(r)dt^2+\frac{dr^2}{f(r)} 
+r^2(\chi'(r)^2dr^2 +\sin^2\chi(r)
d\Omega^2_2)\right]~,
\end{equation}}
\hspace{-.3cm} where $t,r,(\theta,\phi)\in S^2$ are the world-volume
coordinates which coincide with spacetime coordinates. The Born-Infeld
action is $S=\frac{2N}{\pi}\int dtdr L$, where
\begin{equation}\label{bi} L=
r^2\sin^2\chi\left[ \sqrt{1+r^2f(r)(\chi')^2-\frac{4\pi^2}{\lambda}F^2}-r^2\chi'\right]~.
\end{equation}
The first term is $\sqrt{\det (g + F)}$, with $g_{ab}$ the
worldvolume metric and $F_{ab}$ is the worldvolume electromagnetic
field.  The last term is the Chern-Simons term. Consistent with 
symmetries and equations of motion, one can take $F_{ab}$ with one
nonzero component $F_{rt} = A'_t(r)$.  The canonical momentum $\Pi =
\partial L / \partial A'_t(r)$ is a constant equal to the number of
units of electric flux, k.  Solving for F and substituting into the
equation of motion for $\chi$ yields
\begin{eqnarray}
2 r^4 \sin^3 \chi \cos \chi {\sqrt{1+r^2 f (\chi')^2} \over \sqrt{r^4
    \sin^4\chi + \kappa^2}} ~-~4 r^3 \sin^2 \chi \nonumber \\=
\frac{d}{dr}\left( \frac{r^2f\chi' \sqrt{r^4 \sin^4 \chi +
\kappa^2}}{\sqrt{1+r^2f (\chi')^2}}\right) ~.\label{eqmo}
\end{eqnarray}
We fix the boundary condition at $r\to\infty$ to match the zero
temperature $\frac{1}{2}$-BPS D3-brane.  It is a solution of the same
equation with $f(r)=r^2$ and where, to get the Poincar\'e coordinates,
$(\sin\chi,\cos\chi)$ are replaced by $(\chi,1)$.  Then,
$\chi(r)=\frac{\kappa}{r}$ is an exact solution of (\ref{eqmo}) and
the brane geometry is a simple direct product of $AdS_2$ with radius
$\sqrt{1+\kappa^2}$ and $S^2$ with radius $\kappa$
\cite{Drukker:2005kx}. We seek solutions with the asymptotic behavior
$\chi(r)\sim \frac{\kappa}{r}$ for large $r$.  By studying the large
$r$ regime, it is easy to see that there is no solution of
(\ref{eqmo}) for $\chi$ which goes to zero at least as fast as
$r^{-1}$ unless $\kappa$ is non-zero. By studying the region near the
horizon, we can see that there is no solution in the large $\kappa$
limit.  So, if there is a solution at all, it will only exist if kappa
is non-zero but not large.  We have attempted to solve (\ref{eqmo})
numericaly with small values of $\kappa$. 
We have positive evidence for
a solution in a corner of the parameter space obtained by
taking the infinite temperature limit (replacing $r$ with $rL$, $r_+$
by $r_+L$, $\chi$ with $\chi/L$ and taking $L\to \infty$).  The
resulting differential equation has an exact solution for $r_+ = 0$,
$\hat\chi = \kappa /(r+b ) + {\cal{O}}(r_+^4)$ where $b$ is an
integration constant.  Restoring $r_+ > 0$, we employed a shooting
technique to look for solutions which asymptote to $\hat\chi$.  With
$\kappa = 0.001$, there appears to be a solution at
$b\sim 10^{10}$.  Our work is on-going and we shall present the
details elsewhere.



\noindent
The authors acknowledge hospitality of the Galileo Galilei Institute,
Aspen Center for Physics and Perimeter Institute.  This work is
supported in part by NSERC of Canada and the INFN of Italy.


\begin{thebibliography}{99}

      \bibitem{Hawking:1982dh}
        S.~W.~Hawking and D.~N.~Page,
        Commun.\ Math.\ Phys.\  {\bf 87}, 577 (1983).

\bibitem{Witten:1998zw}
  E.~Witten,
  Adv.\ Theor.\ Math.\ Phys.\  {\bf 2}, 505 (1998)
  [arXiv:hep-th/9803131].




\bibitem{Sundborg:1999ue}
  B.~Sundborg,
  Nucl.\ Phys.\  B {\bf 573}, 349 (2000)
  [arXiv:hep-th/9908001].


\bibitem{Polyakov:2001af}
  A.~M.~Polyakov,
  Int.\ J.\ Mod.\ Phys.\  A {\bf 17S1}, 119 (2002)
  [arXiv:hep-th/0110196].



\bibitem{Aharony:2003sx}
  O.~Aharony, J.~Marsano, S.~Minwalla, K.~Papadodimas and M.~Van Raamsdonk,
  Adv.\ Theor.\ Math.\ Phys.\  {\bf 8}, 603 (2004)
  [arXiv:hep-th/0310285].



\bibitem{Polyakov:1978vu}
  A.~M.~Polyakov,
  Phys.\ Lett.\  B {\bf 72}, 477 (1978).


\bibitem{Susskind:1979up}
  L.~Susskind,
  Phys.\ Rev.\  D {\bf 20}, 2610 (1979).


\bibitem{McLerran:1980pk}
  L.~D.~McLerran and B.~Svetitsky,
  Phys.\ Lett.\  B {\bf 98}, 195 (1981);
  B.~Svetitsky and L.~G.~Yaffe,
  Nucl.\ Phys.\  B {\bf 210}, 423 (1982);
  L.~G.~Yaffe and B.~Svetitsky,
  Phys.\ Rev.\  D {\bf 26}, 963 (1982);
  A.~Dumitru, Y.~Hatta, J.~Lenaghan, K.~Orginos and R.~D.~Pisarski,
  Phys.\ Rev.\  D {\bf 70}, 034511 (2004)
  [arXiv:hep-th/0311223].



\bibitem{Maldacena:1998im}
  J.~M.~Maldacena,
  Phys.\ Rev.\ Lett.\  {\bf 80}, 4859 (1998)
  [arXiv:hep-th/9803002].










\bibitem{Drukker:2005kx}
  N.~Drukker and B.~Fiol,
  JHEP {\bf 0502}, 010 (2005)
  [arXiv:hep-th/0501109].


\bibitem{Yamaguchi:2006te}
  S.~Yamaguchi,
  Int.\ J.\ Mod.\ Phys.\  A {\bf 22}, 1353 (2007)
  [arXiv:hep-th/0601089];
  S.~Yamaguchi,
  JHEP {\bf 0605}, 037 (2006)
  [arXiv:hep-th/0603208].

\bibitem{Gomis:2006sb}
  J.~Gomis and F.~Passerini,
  JHEP {\bf 0608}, 074 (2006)
  [arXiv:hep-th/0604007];
  J.~Gomis and F.~Passerini,
  JHEP {\bf 0701}, 097 (2007)
  [arXiv:hep-th/0612022];
  J.~Gomis, S.~Matsuura, T.~Okuda and D.~Trancanelli,
  JHEP {\bf 0808}, 068 (2008)
  [arXiv:0807.3330 [hep-th]].



\bibitem{RodriguezGomez:2006zz}
  D.~Rodriguez-Gomez,
  Nucl.\ Phys.\  B {\bf 752}, 316 (2006)
  [arXiv:hep-th/0604031];
  K.~Okuyama and G.~W.~Semenoff,
  JHEP {\bf 0606}, 057 (2006)
  [arXiv:hep-th/0604209];
  S.~Giombi, R.~Ricci and D.~Trancanelli,
  JHEP {\bf 0610}, 045 (2006)
  [arXiv:hep-th/0608077].



\bibitem{Erickson:2000af}
  J.~K.~Erickson, G.~W.~Semenoff and K.~Zarembo,
  Nucl.\ Phys.\  B {\bf 582}, 155 (2000)
  [arXiv:hep-th/0003055];
  V.~Pestun,
  arXiv:0712.2824 [hep-th].


\bibitem{Hartnoll:2006hr}
  S.~A.~Hartnoll and S.~Prem Kumar,
  Phys.\ Rev.\  D {\bf 74}, 026001 (2006)
  [arXiv:hep-th/0603190].

\bibitem{Gross:1980he}
  D.~J.~Gross and E.~Witten,
  Phys.\ Rev.\  D {\bf 21} (1980) 446.


\bibitem{Jurkiewicz:1982iz}
  J.~Jurkiewicz and K.~Zalewski,
  Nucl.\ Phys.\  B {\bf 220}, 167 (1983).


\end{thebibliography}
\end{document}